\documentclass[11pt,twoside]{article}
\usepackage{asp2010}

\resetcounters

\begin{document}

\title{SAMI Automated Plug Plate Configuration}

\author{Nuria~P.~F.~Lorente, Tony~Farrell, and Michael~Goodwin
\affil{Australian Astronomical Observatory, PO Box 915,\\
North Ryde NSW 1670, Australia}
}

\begin{abstract} The Sydney-AAO Multi-object Integral field
  spectrograph (SAMI) is a prototype wide-field system at the
  Anglo-Australian Telescope (AAT) which uses a plug-plate to mount
  its $13 \times 61$-core imaging fibre bundles (hexabundles) in the
  optical path at the telescope's prime focus.
  In this paper we describe the process of determining the positions
  of the plug-plate holes, where plates contain three or more stacked
  observation configurations. The process, which up until now has
  involved several separate processes and has required significant
  manual configuration and checking, is now being automated to
  increase efficiency and reduce error. This is carried out by means
  of a thin Java controller layer which drives the configuration
  cycle. This layer controls the user interface and the C++ algorithm
  layer where the plate configuration and optimisation is carried
  out. Additionally, through the Aladin display package, it provides
  visualisation and facilitates user verification of the resulting
  plates. \end{abstract}

\section{Introduction}

The Sydney-AAO Multi-object Integral Field Spectrograph (SAMI) is a
prototype wide-field system at the Anglo-Australian Telescope (AAT)
deploying $13 \times 61$-core imaging fibre bundles (hexabundles) over
a 1-degree field of view \citep{2012clb+}. The hexabundles, together
with ancillary sky and calibration fibres, are mounted on a plug plate
located at the prime focus of the telescope. Each plate is a 24~cm
diameter and 3~mm thick steel disc, pre-drilled with holes
corresponding to the on-sky positions of targets for several distinct
pointings. Typically this is 3 fields for science plates, and 8 fields
in the case of plates used for the set-up and calibration of the
instrument.
A science field consists of a guide-star located at the centre of the
plate, a flux calibration star, 26 blank-sky positions and 12 science
targets located in the 1-degree field-of-view. A calibration field
contains either 2 visual alignment stars, used in the initial coarse
plate rotation alignment process, or 13 stars used for distortion
calculations and fine plate rotation alignment.

\section{Configuring the Plate}
\begin{figure}[!t]
\centering
\includegraphics[width=0.7\linewidth]{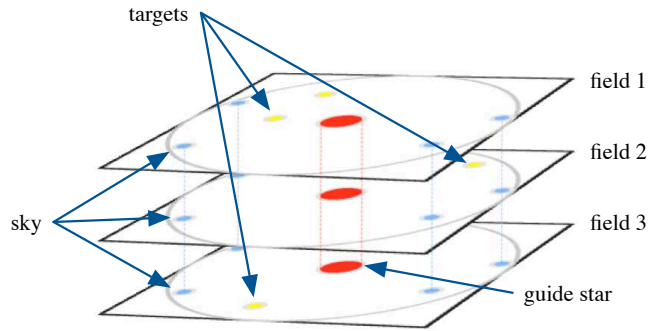}
\caption {Representation of a science plate composed of three fields. The target
positions are unique to a given field, whereas the sky positions are
valid for each of the three field but use a single common hole on the plate.}
\label {p048_FigStackedFields}
\end{figure}

The process of determining the positions of the science plate holes
involves defining 3 stacked observing fields composed of a common
guide-star hole at the centre of the plate, and choosing the flux star
and science targets for each field taking into account separation
constraints between targets in the same field and in the two other
fields (simplistically, plug-holes should not overlap). The 26
blank-sky positions are allocated to each field with the additional
constraint that only 26 sky holes are to be drilled in the plate, so
the sky positions for each of the 3 fields must map to the same
physical holes (see Figure~\ref{p048_FigStackedFields}).
Configuration of the calibration plate holes is simpler, as there are
no sky positions to find which are common to all fields, but star
plates contain a greater number of stacked fields with one common
guide-star hole: 4 visual alignment fields plus 4 fields for the
astrometric distortion model and fine rotation alignment.

\noindent The steps involved in configuring a plate are as follows:
\begin{enumerate}
\item {\bf Field Creation:} From a target catalogue and a catalogue of
  suitable calibration and guide stars, a database of candidate
  observing fields is constructed, based on several positional
  constraints set by the physical limits imposed by the fibre
  connector assembly, the position of the guide camera and the need to
  physically place and remove connectors between one observation and
  the next. \newline 
- Separation between a target and the guide star: $7.6$\ arcmin
$<$~separation $<0.5$~degrees\\
- Separation between two targets in the same field $> 3.8$\ arcmin

\item{\bf Field Stacking:} Taking into account inter-field exclusion
  rules, where the separation of two targets in different fields
  $>$3\ arcmin, multiple fields are stacked to form a single plate.
  Each field is taken from a different RA range, so that a night’s
  observation of 3 fields can be performed without the need to stop to
  change plates.

\item{\bf Sky Fibre Positioning:} A new method is being implemented
  for determining the position of the 26 sky fibres
  (Figure~\ref{p048_FigSkyPos}). First a grid is placed over the field
  with cell size = hole exclusion size / 2.  Any cells which fall within the
  exclusion zone of any star or target are then removed from the grid.
  The most isolated remaining cell is then located (i.e. the cell
  furthest from its nearest neighbour) and tested for suitability as a
  sky position. This involves checking that, in each of the 3 fields
  comprising the plate there are no sources above the magnitude limit
  within the sky fibre's field-of-view.  If the position is suitable,
  the grid cell is marked as full and the nearest-neighbour distances
  are recalculated.  This process is iterated until all 26 sky
  positions have been obtained.

\begin{figure}[ht]
\centering
\includegraphics[width=0.7\linewidth]{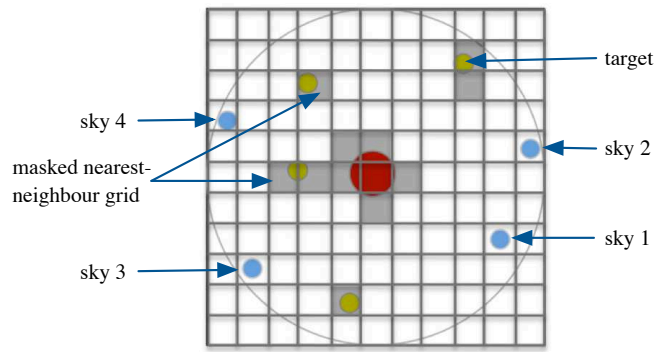}
\caption {Finding suitable positions for sky fibres to ensure
  uniform coverage over the plate. This example shows the invalid
  (i.e. occupied) grid cells as gray and the order in which the sky
  positions have been determined (1 - 4).}
\label {p048_FigSkyPos}
\end{figure}

\item{\bf Transformation:} Once the plate positions are defined,
  instrument and sky compensation transforms are applied. These
  include the optical distortion model and the atmospheric
  differential refraction models, based on the expected time of
  observation of each field

\item{\bf Generation:} The configuration process creates two products:
  the first is a CSV file containing the hole positions for the plate. A final
  adjustment is made for the estimated dome temperature at observing time
  and ambient temperature at fabrication (when the holes are drilled)
  and these data are sent to the plate manufacturer. The second
  product is a set of observing files, which are used to drive the Telescope
  Control System at observing time.

\item{\bf Verification and Visualisation:} Visualisation is done using
  the Aladin application \citep{2000bfb+}. The various components
  making up a plate --- the fields-of-view of the telescope,
  hexabundles, sky fibres and guide camera, the hole information and
  exclusion zones are separately layered onto an image of the
  field (Figure~\ref{p048_FigAladin}). This is done using the Aladin FoV mechanism, sending the
  relative positions from the centre of the field in a VOTable
  format \citep{2011owd+}. The survey target positions are also plotted as a separate
  absolute position $(\alpha, \delta)$ layer. This facilitates
  verification by eye of several items, including:
1) The target position is within the field-of-view of the fibre
  hexabundle and that both of these coincide with the DSS galaxy;
2) The field-of-view of the sky fibres is empty, as expected;
3) The distribution of sky fibres is uniform across the plate; and
4) There are no bright stars or extended galaxies sufficiently
  close to a target position so as to pose a contamination risk.


\begin{figure}[ht]
\centering
\includegraphics[width=0.73\linewidth]{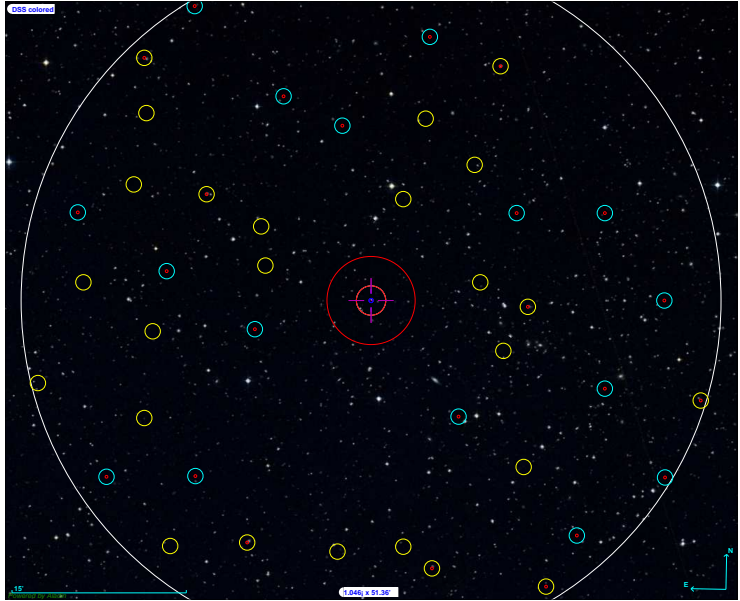}
\caption{Aladin screenshot showing SAMI's $1^\circ$ field-of-view (white),
  guide camera (red), hexabundle (yellow)
  and sky fibre (blue) positions, superimposed on a DSS image.}
\label{p048_FigAladin}
\end{figure}

\end{enumerate}

\section{Automation}
As part of a prototype project this configuration process has until
now been a painstaking task involving the use of several software
packages, scripts, stand-alone code and a lot of manual configuration
and checking. As SAMI evolves from a technology demonstrator to a
survey instrument with an expected observing catalogue of several
thousand targets, this approach will no longer be feasible, for
reasons of both efficiency and the increased likelihood of error.

Consequently, the process for configuring SAMI plates is now in the
process of being automated. This consists of a C++ layer which carries
out the optimisation of target and sky positions for each field and
plate, and applies the required atmospheric, telescope and thermal
models to convert between sky and plate positions. This process is
controlled by a Java layer which also provides visualisation of the
process to the user by means of Aladin, using
VOTables as the data transport mechanism. The aim is to take away the
tedium of plate configuration, whilst giving the user control over the
process, by presenting them with a way of easily checking the validity
of an automatically generated plate and allowing them to drive
subsequently finer configuration cycles until a satisfactory plate
configuration is achieved.

\bibliography{P048}

\end{document}